\documentclass[twocolumn,english,aps,floatfix,pra]{revtex4-1}
\usepackage{amssymb}
\usepackage{graphicx}
\usepackage{amsmath}
\usepackage{pstricks}
\usepackage{ifthen}

\usepackage[percent]{overpic}
\usepackage{tikz}

\DeclareMathSizes{7}{7}{5}{4} 
\DeclareMathSizes{6}{6}{4}{3} 
\DeclareMathSizes{5}{5}{3}{2} 

\DeclareMathSizes{4}{4}{2}{1}

\usepackage{lmodern}
\usepackage{ogonek}

\begin{document}
\title{Classical field records of a quantum system: their internal consistency and accuracy
 }

\author{Joanna Pietraszewicz$^{1}$, Piotr Deuar$^{1}$}
\affiliation{
\mbox{$^1$Institute of Physics, Polish Academy of Sciences,
Aleja Lotnik\'{o}w 32/46, 02-668 Warsaw, Poland}
}
\date{\today}

\begin{abstract}

 We determine the regime where the widespread classical field description
  for quantum Bose gases is quantitatively accurate 
 in 1$d$, 2$d$, and 3$d$ by a careful study of the ideal gas limit. 
Numerical benchmarking in 1D shows that the ideal gas results carry over
 unchanged into the weakly interacting gas. 
 The optimum high energy cutoff is in general shown to depend strongly on the observable in question
  (e.g. energy, density fluctuations, phase coherence length, condensate fraction). 
 This explains the wide spread of past results. 
  A consistent classical field representation with less than 10\% deviation
  in all typical observables can be given for systems at temperatures
   below $0.0064$ degeneracy temperature in 1$d$,
  and $0.49$ critical temperature in 3$d$. 
  Surprisingly, this is not possible for the 2$d$ ideal gas
  even at zero temperature because mean density,
  density fluctuations and energy cannot be simultaneously
   matched to the quantum results. 

\end{abstract}

\maketitle

\section{Introduction}
\label{INTRO}

  The quantum mechanics of a wide variety of physical systems
  can be quite accurately described by an appropriately chosen 
  ensemble of complex fields (also called,
  classical or c-fields) \cite{Brewczyk07,Blakie08,Proukakis08,book12,Kagan97}. 
  Examples include quantum gases of ultracold atoms,
  coherent light fields,  and solid state polariton systems. 
  A common feature is the appearance of collective behavior such as
  high amplitude phase fluctuations and superfluid defects,
  that strongly fluctuate away from the mean field.
  Though the term ``classical'' is used, we are talking about
  the opposite regime to the usual gas of classical {\it particles}. 
  Here it is the collective {\it field}
  that has classical properties such that each member of the ensemble could be 
  non-destructively tracked, while the particles lose their individual
  identity.
  Examples of such approaches include classical field ensembles
  \cite{Goral01,Davis01,Berloff02,Brewczyk07,Blakie08},
  the stochastic Gross-Pitaevski equation \cite{Stoof99,Gardiner03,Proukakis08},
  the truncated Wigner representation \cite{Steel98, Sinatra02, Polkovnikov10} for ultracold atoms, 
  and the open stochastic classical field equations for polaritons \cite{Wouters09,Chiochetta14}. 
  Related approaches for fermions include stochastic mean field theory
  in e.g. heavy ion collisions\cite{Ayik08,Lacroix13},
  and effective field theories for the pairing order parameter\cite{Klimin15,Simonucci14}.

  In the absence of sufficient {\sl in situ} experimental resolution,
  the approach is also commonly used like a flight recorder
  to give information on the dynamics of the system
  before its detection in destructive 
  time-of-flight images. Its applications are growing in importance given
  advances in the experimental investigation of spontaneous
  superfluid defects and phase fluctuations, such as
  \cite{Gring12,Langen13,Chomaz15,Lamporesi13,Donadello14,Serafini15,Navon15,Sadler06,Weiler08,Parker13}. 
  In quantum many-body systems with collective nonlinear phenomena,
  such ensembles of complex fields are often the only
  practical way to obtain theoretical information on fluctuations, full distribution functions, 
  and -- especially -- on typical single realizations with superfluid defects or
  quasicondensate phase fluctuations \cite{Dettmer01, Berloff02, Davis02, Witkowska11, Barnett11, Karpiuk12, Nowak12, Schmidt12, Weiler08, Sabatini12, Simula14, Bisset09, Witkowska14, Witkowska13, Swislocki13, Martin10a, Martin10b, Cockburn11b, Liu14}.

  However, their use has usually been accompanied by lingering doubt on whether
  the results are quantitative or qualitative. 
  From an operational perspective, two major contributing factors
  to that have been (1) a visible dependence of some results on
  the high energy cutoff that is chosen, and
  (2) different optimum cutoff values found in the literature 
\cite{Witkowska09,Sinatra02,Cockburn11,Brewczyk07,book12,Castin,Bradley05,
Karpiuk10,Davis03,Zawitkowski04,Davis02,Brewczyk04}.
  The aim of this paper is to identify a regime where c-fields are in fact quantitatively accurate, so that they can be used there with confidence.

  Qualitatively, the condition for the applicability of
  classical fields to bosons is that the relevant
  physics can be captured by considering only the highly 
  occupied single-particle modes, without the need for a condensate \cite{Kagan97}. 
  Poorly occupied modes are not described well, and those above an energy cutoff need to be discarded
  to avoid pathological behavior such as the UV catastrophe known since late 19th century physics. 

  However,
  the matter of just where to draw the line and how accurate
  the description is, has been a matter of much contention and ambiguity.
  The history of applying classical fields to ultra cold atomic
  gases teaches us that accuracy has depended
  quite strongly on the choice of the high energy cutoff and the observables studied.
  Past numerical benchmarking
  \cite{Sinatra02,Bradley05,Witkowska09,Karpiuk10,Cockburn11,Bienias11,Witkowska10,Bienias11b,Karpiuk12,Wright11,book12,Blakie08a,Davis03}, 
   careful comparisons to experiment \cite{Cockburn11c,Cockburn11,Galucci12,Cockburn12}, 
  and also analytical \cite{Castin,Witkowska09}
  and purely mathematical studies \cite{Lewin15}
  of various single observables have found that
  it is possible to achieve good to very good agreement,
  but the details of the recipe vary from study to study
  \cite{Sinatra02,Davis03,Brewczyk07,Witkowska09,Cockburn11}.

  Here, we intend to clarify these dependencies,
  and will show that under the right conditions
  the classical field approximation can be treated as more
  than just a qualitative guide, but 
  gives predictions that are correct within small error bounds for a wide range of observables. 

  We will concentrate first on the case of an ideal gas as a baseline,
  reasoning that well described interacting regimes can be found
  at temperatures that are already well-described in the ideal gas.
  Then we will confirm that accuracy seen in the ideal gas
  carries over into the weakly interacting regime under appropriate conditions.
  We will work in the local density approximation (LDA) in the thermodynamic limit.
  That is, we will consider pieces of the gas cloud 
  having a certain \emph{local} density, which allows us to remain general in terms of trap geometry. 
  In the LDA, it is natural to work
  in the grand canonical ensemble (GCE),
  where the rest of the system acts
  as the particle and thermal reservoir. 
  Such a model underpins more general behavior,
  and it will be seen that several important conclusions can be reached. 
  
  In Sec.\ref{APPR} we will describe our approach.
  Further, in Sec.\ref{EIGEN}, we will  
  find the temperature dependent ``eigen'' cutoffs that allow
  the classical fields to correctly match the density and
  one other observable.
  Subsequently, in Sections \ref{rms} and \ref{RMS}
  we will determine the resulting errors
  in other observables and the cutoff that minimizes
  the systematic error across the whole range of observables. 
  This will tell us about the temperature range
   over which an accurate complex field description of the system
  is possible. Finally, via numerical benchmarking of a weakly interacting gas in 1$d$
  to the exact Yang and Yang solution \cite{Yang69}
  we will show  in Sec.\ref{INTER} 
  that the ideal gas results carry over largely unchanged into that regime.
  We conclude in Sec.\ref{CONC}.

\section{Approach}
\label{APPR}

\subsection{Classical field description}
\label{CFIELD}

 The essence of the classical fields method is to replace   
 annihilation (creation) operators  $\hat{a}_k$ ($\hat{a}_k^{\dagger}$)
 of single particle modes in the field operator 
 by complex amplitudes~  $\xi_k$~ ($\xi^{*}_k$), 
 which is warranted when occupation is macroscopic. 
 Then we can write: 
 \begin{equation}
       \hat{\Psi}( {\bf x}) = \sum_{k} \hat{a}_k  \psi_k( {\bf x} ) 
       \to \Bigg{\{} \sum_{k\in \mathcal{C}}  \xi_k  \psi_k( {\bf x}) \Bigg{\}} 
 \end{equation}
 where $\psi_k({\bf x})$ is the wave function for the $k$th mode
 and $\mathcal{C}$ denotes the low energy subspace.
 Since we will be considering uniform sections of the gas,  
 plane wave modes $k\equiv{\bf k}$ are the most convenient,
 with momentum cutoff $k_c$ so that only modes $|\mathbf{k}|<k_c$ are included in $\mathcal{C}$.

 In general, it should be understood that $\hat{\Psi}({\bf x})$ corresponds to 
 an ensemble $\{ ... \}$ of complex field realizations,
 each with its own set of  amplitudes $\xi_k$. 
 The full ensemble preserves the gauge symmetry of the
 quantum thermal state that corresponds to
 a set of many experimental realizations. This is despite the ''virtual" 
 symmetry breaking done by each member of the
 ensemble similarly to a single experimental realization \cite{Kagan97}.

\subsection{Parameters}

   The properties of the uniform dilute gas 
   can be encapsulated by two dimensionless parameters.
   The first is $ \gamma=\frac{m g}{\hbar^2 n} $ 
   with density $n$ and contact interaction strength $g$, 
   and the second is a
   reduced temperature~ $\tau$ which depends on the density $n$, but 
   not on the interaction strength.    
   We choose the thermal de Broglie wavelength
   $\Lambda_T=\sqrt{\frac{2 \pi \hbar^2}{m k_BT}}$ as our length scale, 
   so that the reduced temperature is 
   \begin{equation}
       \tau=\frac{T}{T_{d}} = \frac{1}{2 \pi} \frac{m k_B }{\hbar^2 } \frac{ T}{n^{ 2/d }}.
       \label{eq_tau}
  \end{equation}
   Here, $T_{d}$ is the usual quantum degeneracy temperature
   in $d$ dimensions that corresponds to one particle per region
   of volume $\Lambda_T^d$.
   It is a natural scale for our investigation because then $\tau=1$ corresponds 
   to the point at which the highest mode occupation is $\mathcal{O}(1)$,
   and this constitutes the intuitive ultimate upper bound on temperature
   for which classical field descriptions make sense.

   It is convenient to also scale the cutoff in these units: 
   \begin{equation}
       f_c = k_c \ \frac{\Lambda_T}{2\pi}.
       \label{eq_fc}
   \end{equation}
   A value of $f_c=1$ corresponds to a cutoff at the plane waves with thermal de Broglie wavelength $\Lambda_T$. 
   We will henceforth work in the following units:   
   $\Lambda_T=1$ and $\hbar = m = 1$, where $m$ is the mass of particles. 
   Note that the cutoff in terms of single particle energy is 
  \begin{equation}\label{epsc}
     \varepsilon_c = \pi\,k_BT\,f_c^2.
   \end{equation}

   In the ideal gas limit ($\gamma \to 0$) that we consider first,
   there is only one physical parameter characterizing the system --
   the density-dependent reduced temperature $\tau$, and 
   one technical parameter $f_c$ for the classical fields description.
   Phase space density equal to one occurs at 
   $\tau=\tau_D= \{1.539, 1.443, 1.368 \}$
   in 1$d$, 2$d$ and 3$d$ respectively, while the BEC critical temperature in $3d$ is $\tau=\tau_C=0.5272$.     

\subsection{Observables}
\label{OBS}
     
 The great majority of experiments 
 concentrate on low order observables
 such as phase, density or their fluctuations. 
 We will analyze the following:
 
 1.\ $n$ -- density.
 
 2.\ $\varepsilon$ -- kinetic energy per particle.
 
 3.\ $l_{pg}$ -- phase grain length. \\
  This is the size of a coherent region,
  which we will calculate~ via~
  $l_{pg}:=\frac{1}{n}~\int d{\bf z} \,
  {\big\langle} \hat{\Psi}^{\rm \dagger}(0)\hat{\Psi}({\bf z}) {\big\rangle}
   = \int d{\bf z}\, g^{{\rm (1)}}({\bf z})$.
  In the quasicondensate regime, when 
  $g^{\rm (1)}({\bf z})\simeq e^{-|{\bf z}|/l_{\phi}}$,
  $l_{pg}$~ equals the phase coherence~ length~ $l_{\phi}$.
  
  4.\ $g^{\rm {(2)}}(0)$ -- normalized local density fluctuations. \\
  While these are of much theoretical interest,
  they are rarely measured \emph{in situ}
  because imaging resolution is usually much worse
  than the intrinsic density correlation length of the system. 

  5.\ $u_G$ -- coarse-grained density fluctuations.\\  
  This quantity is defined as
  $u_G := {\rm var} N /\langle N \rangle =
  n\, \int d{\bf z}\, {\big[} g^{\rm (2)}({\bf z})-1 {\big]} + 1$, where
  $N$ is the atom number in a region much larger 
  than  the density correlation length. 
  In contrast to $g^{(2)}(0)$, this intensive thermodynamic quantity often appears in
  experimental work \cite{Armijo11,Armijo} and 
  it gives the ratio of the measured fluctuations in a pixel 
  to Poisson shot noise.  It is equal to the static structure factor at $|\mathbf{k}|=0$, i.e. $S(0)$ .

   6.\ $\rho_o$ -- condensate fraction.

   7.\ $a_r$ -- coherence half width. \\
   In the presence of a true condensate,  $l_{pg}$ (and $u_G$)
   ceases
   to be a good thermodynamic quantity, diverging 
   because $g^{\rm (1)}(z\to \infty) = \rho_o$. In light of this we 
   need another measure of the width of phase fluctuations, 
   and will define it by the half width of the peak of $g^{(1)}(z)$,
   i.e.  $g^{\rm (1)}(a_r)= \frac{1}{2} \big(1 + \rho_o\big)$.
 
  It is worth noting that the kinetic energy per particle
  in itself is not a typical subject of measurement,
  but its consideration has here its own justifications.
  If typical observable quantities are described correctly, but
  $\epsilon$ is not, then this will quickly come out as errors in the dynamics.

\subsection{Ensemble}
\label{ENS}

 A major consideration in our work here has been to remain
 independent of trap geometry. 
 This basically requires working in the local density approximation (LDA). 
 As an example of variations with geometry that can occur without an LDA approach,
 optimal energy cutoffs found on the basis of the distribution of condensate fraction
 for a whole cloud in the canonical ensemble were $0.29 k_BT$ in a uniform box,
 but $1.0 k_BT$ for a harmonically trapped gas \cite{Witkowska09}.
 In the end in Sec.~\ref{DISC}A, we will see that the results of the LDA approach
 can be largely reconciled with the harmonically trapped canonical ensemble results. 

 When considering a relatively uniform section of a larger gas,
 it is not only possible, but also essential to work in the grand canonical ensemble (GCE)
 rather than the canonical one.
 In such a situation the rest of the system acts as a particle and thermal reservoir, 
 while the uniform GCE section describes the properties that are local
 to the region. This approximation is acceptable provided the physical length scales such as $l_{pg}$ are shorter
 than the length scale on which the density changes.
  Such conditions generally prevail for quasicondensates
  or a 3$d$ gas above the condensation temperature. 

 Use of the GCE in a truly condensed system such as the 3$d$ gas below $T_c$
 or the finite-size 2$d$ gas at extreme low temperatures,
 requires some care and background to get our bearings.
  It is known that for the ideal gas
 the usual thermodynamic equivalence between ensembles
 is lost in the presence of condensation.
 Particularly glaring differences are seen in the fluctuations of condensate fraction
 between the canonical and grand canonical
 ensemble --- a matter that has been much
 studied \cite{Grossmann96, Politzer96, Gajda97, Navez97, Grossmann97, Giorgini98, Weiss97, Idziaszek99}
 and is sometimes  known as the ``fluctuation catastrophe'' for the GCE.
 In fact, a uniform condensed system in the GCE has anomalous fluctuations
 of the number of condensed  particles
 (i.e. their variance grows faster than the mean number),
 which implies that some quantities such as $u_G$ diverge. 
  Technically this signals the point of the breakdown
  of the theory \cite{Yukalov05,Yukalov09,Yukalov11}, but in reality this kind of behavior
  cannot actually occur. 
  Physically the growth of diverging quantities is braked by
  other effects.
  Usually, the causes can be traced to
  either a breakdown of the thermodynamic limit
  due to finite size effects, or a suppression of fluctuations due to
  interactions (see \cite{Kocharovsky06}  for a detailed discussion).
 
  The primary difference between the grand canonical 
  and microcanonical or canonical treatments of an ideal condensed system
  has been pointed out quite early \cite{Politzer96} by studying
  the ground state number fluctuations. They are huge  in the GCE
  (the occupation $N_0$ of the ground state  
  is exponentially distributed  $P(N_0)\sim e^{\mu N_0/T}$)
  but small in the other thermodynamical ensembles.
  In contrast, there is no such difference
  for excited level occupations.  
  This suggests that  the majority of observables
  are not pathological.
  Even the mean condensate fraction
  does not diverge nor break equivalence between ensembles, unlike its fluctuations. 
  Hence, it is legitimate to benchmark classical fields in the GCE provided
  that we exclude from  consideration those observables that are known to be deviant.
  In particular, when condensation is present 
  neither condensate fraction  $\rho_o$ nor the main coherence
  decay described by $a_r$ are pathological,
  so we will use these instead of $u_G$ and $l_{pg}$.
  
  The suppression of anomalous condensate fluctuations due
  to interactions can occur even at very weak interactions.
  This can be seen from a simple argument:
  Consider the GCE partition function of the condensate mode:
  \begin{equation}\label{Z_0}
       Z_0(\mu,T) = \sum_{N_0=0}^{\infty} e^{(\mu N_0 - C_0 g N_0^2)/T}
  \end{equation}
  with $g$ the interaction strength,
  and $C_0$ a geometry-dependent factor \cite{Politzer96}.
  In this form it is 
  now a Gaussian distribution of the condensate occupation with mean
  $\overline{N}_0 = \mu/2gC_0$ and relative condensate number fluctuations
  $\Delta N_0 / N_0 = \sqrt{T/(2C_0g N_0^2)} = \sqrt{T/(2E_{\rm int})}$.
  It means that 
  the relative magnitude of the number fluctuations is  related to
  the ratio of the temperature to the interaction energy of the \emph{entire} system, $E_{\rm int}$.
  The latter very quickly suppresses the grand canonical fluctuation catastrophe as the size of the system
  becomes appreciable, leaving only a tiny low temperature region
  at $T\lesssim 1/E_{\rm int}$ with anomalous fluctuations, that shrinks as $T \to 0$.
  
  The above considerations are distinct from the separate matter of
  what ensemble should be considered for the \emph{entire} system.
  If one were to nondestructively follow a single realization of the system over time
  and assume ergodic evolution, then the correct ensemble would be the microcanonical one
  that has the system isolated from particle and energy exchange.
  This has been considered in many works \cite{Gajda97,Navez97, Kagan97,Stoof99,Goral02}.
  On the other hand, actual experimental studies usually deal with
  an ensemble over many independent realizations
  created by cooling a new cloud each time,
  and independently measuring each destructively.
  Then the fluctuations of the number of particles between
  different realizations can in fact be 
  of the same order as the mean number of particles over the whole experimental series.
  Due to the large number fluctuations between shots,
  a sequence of single clouds is likely to be more closely described by
  the GCE than the CE.
  
  To wrap up this section,
  once correlation length scales are short enough for the LDA to be valid,
  the approach used here is relevant also to global properties of the system
  when the GCE fluctuation catastrophe is suppressed.
  This can happen because of any of the following:
  (1) lack of a true condensate,
  (2) observables that do not depend on fluctuations of the condensate fraction,
  (3) weak but sufficient interaction for $E_{\rm int}$ of the entire system
  to be large compared to the temperature,
  (4) an experimental data set that consists of an ensemble of many independent realizations,
   except for the cases with  strong post-selection on particle number. 
   Such conditions prevail in a very wide range of systems of interest.
   Keeping these in mind, let us proceed. 

\subsection{Benchmarking}
 
   We will compare the classical field predictions
   for the observables in Sec.~\ref{OBS} to the exact Bose gas values
   in the thermodynamic limit. For the ideal gas they can mostly be obtained analytically.

   To proceed, the LDA approximation requires first the density to be correct,
   in the sense that an ideal gas with density $n$
   (i.e. reduced temperature $\tau$) should be compared to a classical field ensemble
   with the same density. 
   This is also essential in practice regardless of the LDA,
   since $n$ is the most basic observed quantity in experiments.
   To match ideal gas and classical field densities,
   firstly chemical potentials  $\mu^{\rm (id)}$ and $\mu^{\rm (cf)}$, respectively, must be chosen. 
   A sum over Gibbs factors gives the exact Bose gas density
   $n\big(\mu^{\rm (id)}\big)$ and the density estimate
   $n^{(\rm cf)}\big(\mu^{\rm (cf)},f_c\big)$
   in classical fields as functions of their grand canonical chemical potentials. 
   We invert these, and  with the help~ of~ Eq.~(\ref{eq_tau}) 
   obtain $\mu^{\rm (id)}(\tau)$ and $\mu^{\rm (cf)}(\tau,f_c)$. 
   Other observables  e.g. $\varepsilon^{\rm (id)}(\tau)$ and 
   $\varepsilon^{\rm (cf)}(\tau,f_c)$ 
   can then be expressed as functions of $\tau$ and $f_c$ as well.  
   
   In general, for the Bose ideal gas, $\tau$ and the choice of units
   specify all properties of the system.
   In classical fields, in addition to $\tau$,
   the system description 
   requires a technical parameter $f_c$.
   There, we can fit both densities
   $n^{(\rm id)}$ and $n^{(\rm cf)}$ to $\tau$,
   but also we can make one other quantity agree 
    exactly by an appropriate choice of $f_c$.

\section{Observable-dependent accuracy}
\label{ACC}     
\subsection{Single observable ``eigen'' cutoffs}
\label{EIGEN}
\setlength{\unitlength}{1cm}
\begin{figure}[b]      
    \includegraphics[width=8.5cm]{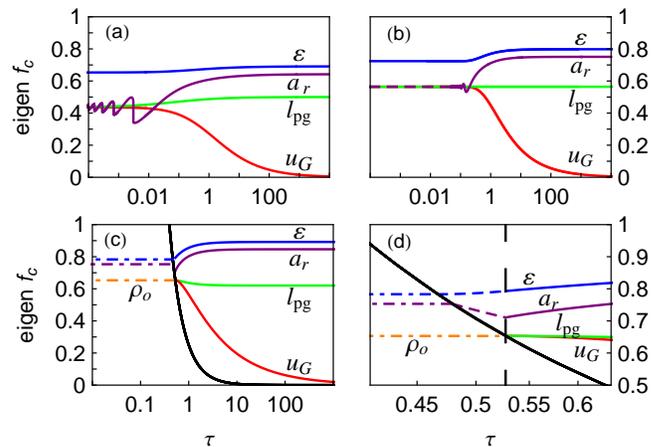}
        \vspace*{-0.2cm}
      \caption{(Color online) Matched eigen cutoffs $f_c$ 
        for several observables as~ a~ function of temperature $\tau$
        ($\varepsilon$ -- blue line, $a_r$ -- purple line,
        $l_{pg}$ -- green line, $u_G$ -- red line, $\rho_o$ -- orange line).  
        The~ top panels (a), (b) show 1$d$ and 2$d$ cases, respectively,
        and the~ bottom panels present the 3$d$ situation
        with (d) a magnification of the critical region. 
        The Bose gas critical temperature $\tau_C$ is 
        marked as a vertical dashed line, while the black solid line
        shows the $f_c$ value
        below which condensation of classical fields occurs.
        }
        \label{eigFC}   
\end{figure}
  Fig.~\ref{eigFC} shows how such cutoffs matched
  to different observables
  (which we will call \textsl{eigen} cutoffs)
  behave as a function of temperature. 

  The density is already matched due to the LDA as explained above, and is not shown.  
  We have also not shown results for $g^{(2)}(0)$ 
  because it is always correctly predicted to be
  $g^{\rm (2)}(0)=2$ for every cutoff in the ideal gas. 
  This property will not hold any more when interactions are present. 
  Indeed, then the local density fluctuations manifest a dependence on cutoff.
  
  The high temperature
  behavior is qualitatively similar
  in all dimensions. 
  The eigen cutoffs matched to energy per particle $f_c^{\varepsilon}$
  and to coherence half width $f_c^{a_r}$
   rise to constant values, while
  the eigen cutoff matched for density fluctuation
  $f_c^{ u_G}$ drops to zero (this will be commented on later in Sec.~\ref{DISC}B). 
  The $f_c^{ l_{ pg}}$ takes intermediate values and is almost constant.
  An unexpected feature is 
  the similar behavior of cutoffs corresponding to
  $a_r$ and $\varepsilon$ rather than the $a_r$ and $l_{pg}$
  that are more related physically.
   
  The crossover to low temperature behavior 
  is around $\tau=1$, as expected. In the low temperature regime, most eigen cutoffs 
  collapse to a common value ($0.436$~ and~ $0.564$, in 1$d$ and 2$d$, respectively), 
  except for $f_c^\varepsilon$ which prefers
  the higher values $0.653$~ and~ $0.724$.
  In 3$d$, the cutoffs at~ $\tau\to0$ are 
  $0.783$, $0.753$, and $0.653$ for $\varepsilon$, $a_r$, and $\rho_o$, respectively. 

   Below critical temperature in 3$d$, the eigen cutoff for condensate fraction has a constant value. 
   This comes about because the critical temperature in classical fields is cutoff dependent,
   $\tau_C^{(\rm cf)}=[4 f_c]^{-2/3}$,
   while in the Bose gas it is
   $\tau_C=[ \zeta(3/2)]^{-2/3}=0.5272$ with $\zeta(3/2)$ the Zeta function.
   The condensate fractions are directly related as 
   $\rho_o^{\rm (id)}=[1-(\tau/\tau_C)^\frac{3}{2}]$
   and $ \rho_o^{\rm (cf)}=[1-(\tau/\tau^{(\rm cf)}_C)^\frac{3}{2}]$.
   Hence, $f_c^{\rho_o}=\frac{\zeta[3/2]}{4} = 0.65309$ makes
   $\rho_o^{\rm (id)}$ and $\rho_o^{\rm (cf)}$ equal for all $\tau\le\tau_C$.

  Two other noteworthy points are that:
  ($\star$) in 2$d$, the eigen cutoff $f_c~=~1/\sqrt{\pi}$
  that gives the correct phase grain length~ $l_{pg}$
  does not depend on temperature, and
  ($\star\star$) the wave-like behavior of $f_c^{a_r}$
  in 1$d$ (as well as in 2$d$), that
  comes from oscillations of $g^{\rm (1)}({\bf z})$ with distance,
  is caused by the sharp cutoff in momentum space
  in classical fields. 
 

\subsection{Relative errors of single observables}
\label{rms}

  Now, how does a non-optimal choice
   of $f_c$ affect the observables,
  and their systematic error? 
  This is very relevant for practical considerations.
  For one thing, in a nonuniform system, 
  when the cutoff is matched in one spatial region,
  it is good to know the sensitivity of results 
  in other regions with a different density 
  on this choice of~ $f_c$.
  Furthermore, we need this information to judge 
  how good the classical fields are
  in describing the system overall.

\setlength{\unitlength}{1cm}
\begin{figure}[t]
       \begin{picture}(8.5,10)
       \put(0.1,7.5){\resizebox{8.4cm}{2.5cm}{
              \includegraphics{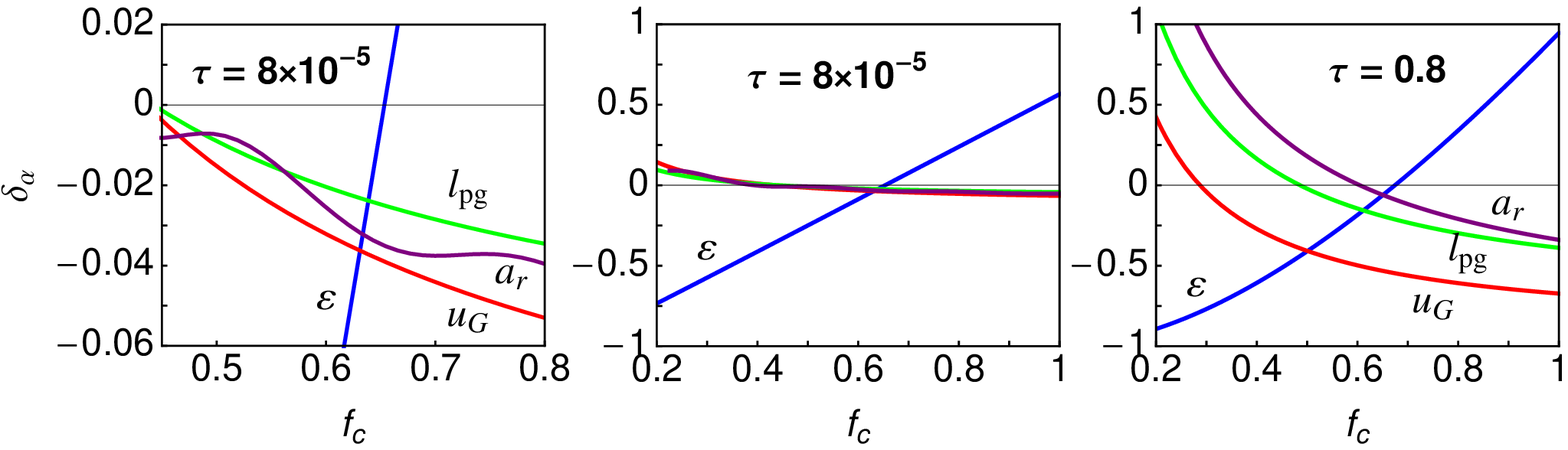}
              }}
       \put(0,10.){\textsf{(1$d$)}}
%
       \put(0.1,5){\resizebox{7cm}{2.45cm}{
              \includegraphics{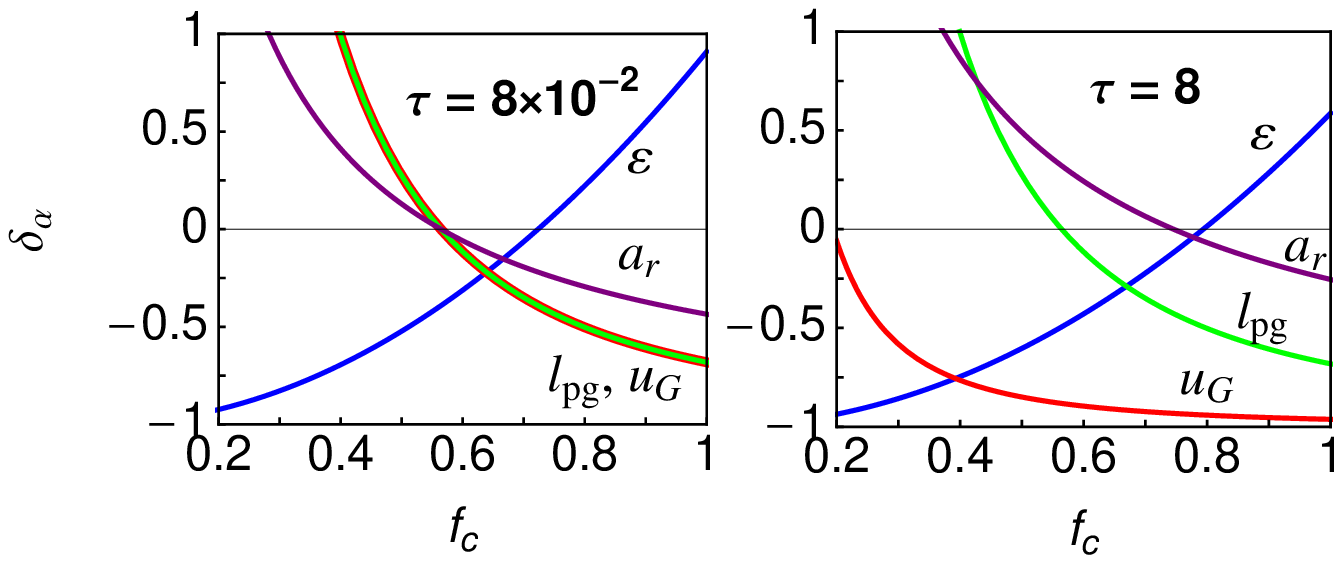}}}
       \put(0,7.4){\textsf{(2$d$)}}
       %
%
       \put(0.1,2.44){\resizebox{8.4cm}{2.5cm}{
              \includegraphics{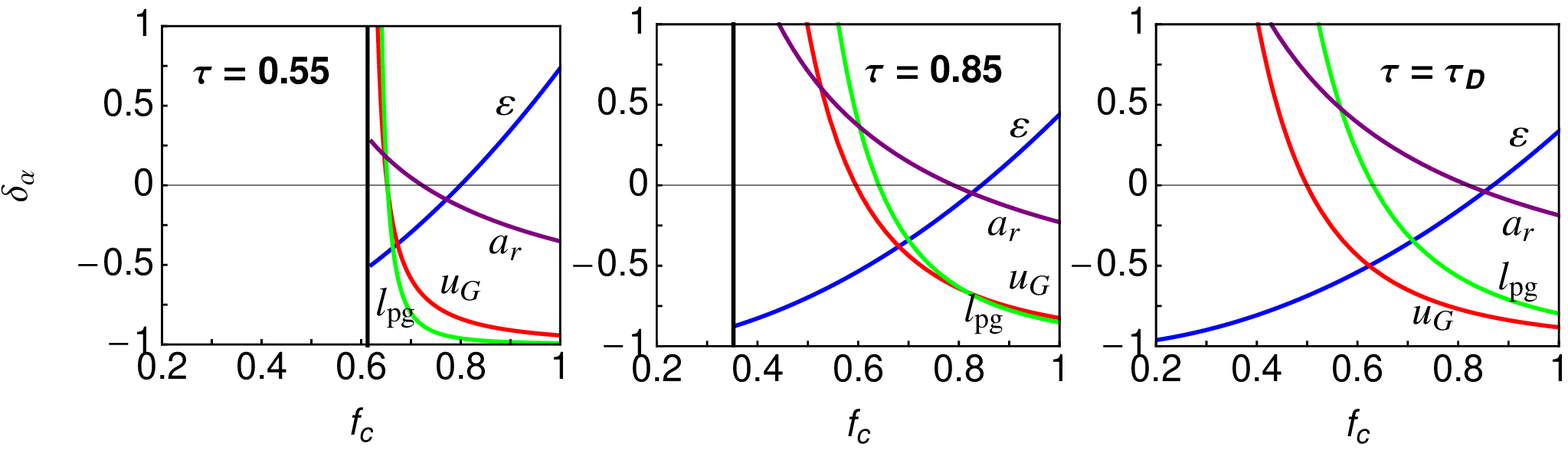}
          }}
          \put(0,4.85){\textsf{(3$d$)}}
       \put(0.1,0){    \resizebox{7cm}{2.45cm}{
              \includegraphics{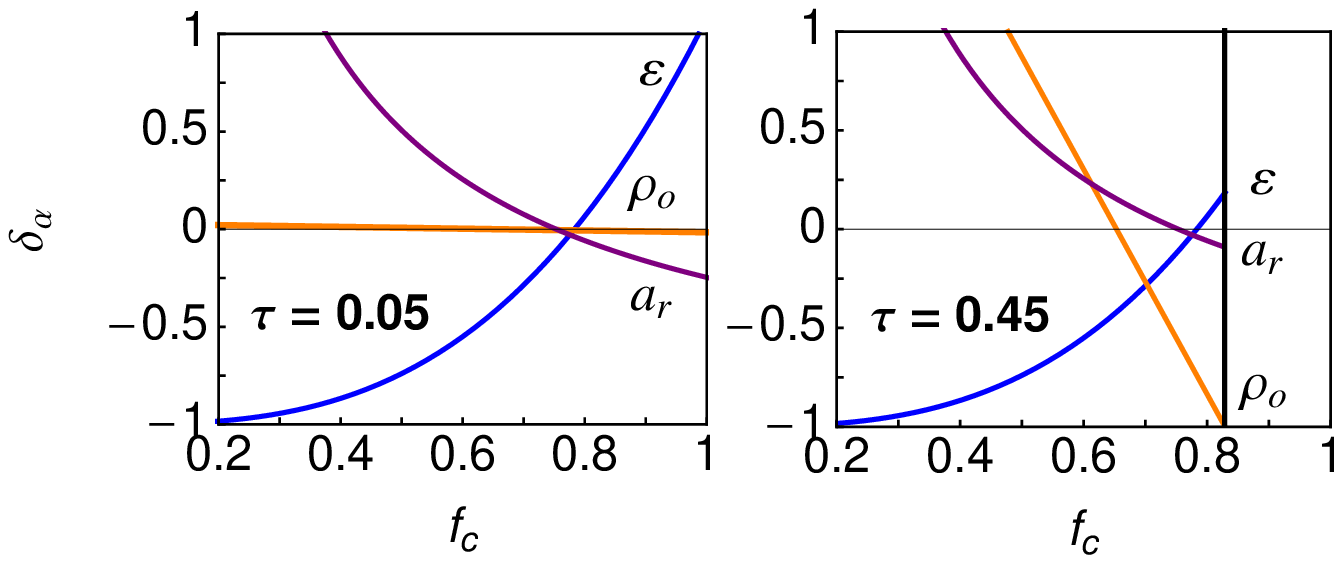}
          }}     
        \end{picture}
        \vspace*{-0.2cm}
      \caption{(Color online) Variation of the relative errors  $\delta_\alpha$ of observables
          with cutoff $f_c$ at representative high and low temperatures $\tau$.
          Colors like in Fig.~\ref{eigFC}.  
          Top row: 1$d$, second row: 2$d$,
          third row: 3$d$ above $\tau_C$, 
          last row: 3$d$ below $\tau_C$. 
          }
       \label{Srms}   
\end{figure}
   The relative error $\delta_\alpha$ of an observable $\alpha$ is:
  \begin{equation}
     \delta_\alpha(\tau,f_c) := \frac{\Delta \alpha}{\alpha} =  
     \Bigg(  \frac{ \alpha^{\rm (cf) }(\tau,f_c) }{ \alpha^{\rm (id) }(\tau)} - 1 \Bigg)  
     \label{malerms}
   \end{equation}
   Its cutoff dependence is shown in Fig.~\ref{Srms}.
   The first observation is that
   the relative error of energy per particle 
   has an opposite trend to 
   the other quantities.
   The resulting mismatch 
   turns out to be the strongest restriction on 
   the $f_c$ range for which all $\delta_{\alpha}$
   errors are small. 
   
   Secondly, in 1$d$ the known fact \cite{Castin} that 
   $g^{\rm (1)}({\bf z})$ and  $g^{\rm (2)}({\bf z})$ do not
   depend on cutoffs at low $\tau$,  
   is reflected in small errors in $l_{pg}$, $u_G$, and $a_r$. 
   However these errors are no longer small in higher dimensions. 
   As temperature drops, the $\delta_{\alpha}(\tau,f_c)$ except for $\delta_{\rho_o}$,
   collapse onto curves
   that stay invariant with $\tau$ and remain steep 
   (the $\tau~=~0.08$ and $\tau~=~0.05$ panels in Fig.~\ref{Srms}). 
	In other words, 
    \textit{observables remain sensitive
    to cutoff all the way down to zero temperature in 2$d$ and 3$d$.}

   
\section{Global accuracy}
\label{RMS}
 What does it take to match 
  all, or at least to be close to all typical observables? 
  Let us consider the global error estimator
  \begin{equation}
       RMS_{\rm { \alpha,\beta,\dots}}(\tau,f_c)= \sqrt{ \Big( \delta_{\alpha} \Big)^2 +
       \Big( \delta_{\beta} \Big)^2 + ...} \nonumber \\
  \end{equation}
  This is a root mean square of the relative errors of 
  chosen observables $\alpha$, $\beta$, etc.
  Each relative error will, by definition,
  be less than $RMS$.
  The main aim of the function $RMS$
  will be to catch inaccuracy in any observable.  
  
  We have studied the $RMS_{\alpha,\beta,\dots}$ with all the observables that we 
  have been considering.
  Moreover, we also took various combinations of them.
  It turns out that when we include just $u_G$ and $\varepsilon$,
  all relevant features that were seen 
  with larger sets of observables are covered.
  This happens because these quantities are the most
  ``extreme'' in terms of the behavior of eigen $f_c$
  and of the values and trends of $\delta_\alpha$. This is
 seen in Figs.~\mbox{\ref{eigFC}--\ref{Srms}}.
  Also, the pair $(\varepsilon,u_G)$ includes
  observables of 2nd and 4th order in $\hat{\Psi}$,
  which are the two main classes measured in experiments.
  We will use them to define the quantity:   
  \begin{equation}
       RMS(\tau,f_c)= \sqrt{ \Big(\ \delta_{ \varepsilon}\ \Big)^2 +
       \Big( \delta_{u_G} \Big)^2 } 
       \label{eqBigrms}
  \end{equation}
  that will be our indicator of the overall accuracy and applicability 
  of the classical fields approximation.
  Below $\tau_C$ in 3$d$, the condensate fraction  $\rho_o$
  will be used instead of $u_G$.
  
  Minimizing Eq.~(\ref{eqBigrms}) at a given temperature  
  will give the optimal cutoff momentum
  and minimum error indicator $minRMS$.
  For example, a $minRMS$ value below $0.1$ (i.e.  $<10\%$ error in observables)
  is often satisfactory and we will take it as a guideline. 
    

\begin{figure}[t]
 \includegraphics[width=8cm,height=4.5cm]{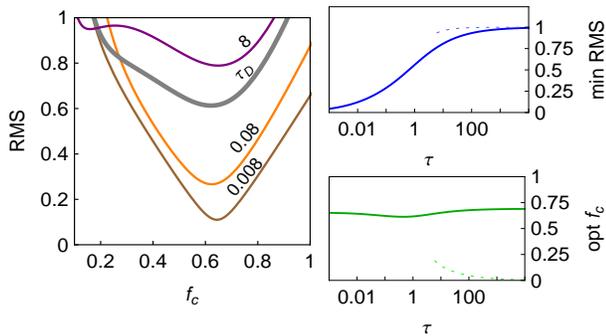}
\vspace*{-0.2cm}
 \caption{(Color online) Summary results for the 1$d$ gas.  
  The left panel shows the dependence of the global error estimator $RMS$,  
  based on $\varepsilon$ and $u_{G}$, 
  on cutoff $f_c$ for several values of $\tau~=~\{ 0.008$ brown,
  $0.08$ orange, $\tau_D$ gray, $8$  purple$\}$.
  The top right panel shows the minimal value of $RMS$ 
  achieved at the optimal cutoff shown in the lower right panel.
  An additional dashed branch indicates a less optimal
  local minimum of $RMS$.
  }
 \label{BigRMSDim1}
\end{figure}
\begin{figure}[b]
\includegraphics[width=8cm,height=4.5cm]{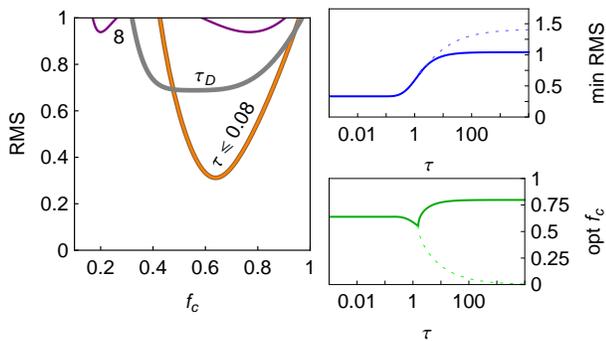}
\vspace*{-0.2cm}
 \caption{(Color online) Summary results for the 2$d$ gas.  
  Description as in~ Fig.~\ref{BigRMSDim1}.
  The $\tau=0.008$ and $\tau=0.08$ lines in the left panel overlap.}
 \label{BigRMSDim2}
\end{figure}
 
  Fig.~\ref{BigRMSDim1} shows the results
  for the 1$d$ gas.
  Global error $RMS$ is very large
  above the degeneracy temperature $\tau=1$.
  For low temperatures it falls to zero,
  as one would hope. According to our $10\%$ guideline, 
  classical fields give acceptable results up to $\tau=0.0064$. 
  The best choice of $f_c$ is fairly
  invariant with temperature in this region,
   being in the range $(0.649\pm 0.043)$.
  In fact, if we choose the average value of $f_c$,
  we will be close to absolute $minRMS$
  regardless of temperature or density.
  At high $\tau$ an extra second branch appears that 
  is associated with a local minimum of $RMS$ with large errors in $\varepsilon$ 
  and small in $u_G$. 
  It is not of practical importance for us because it is less optimal.
  
  Fig.~\ref{BigRMSDim2} shows the results
  for the 2$d$ gas.
  The behavior at low temperature is surprisingly unfavorable.
  $RMS$ never falls below $0.333$.
  This is a consequence of an inability 
  to satisfy both  observables $u_G$ and $\varepsilon$. 
  Their relative errors $\delta_{\varepsilon(u_G)}(\tau,f_c)$
  become stuck 
  on the curves shown in the fourth plot of Fig.~\ref{Srms} whenever $\tau\lesssim 0.08$ 
  and do not cross near zero error.
  One wonders whether this situation ($minRMS$ well above $10\%$ as $\tau\to0$)  
  is repeated for other different sets of observables?
  It turns out that even the pairs 
  $(\varepsilon, l_{pg})$ or $(\varepsilon ,a_{r})$
  will lead to similar large $minRMS$ values.
  In fact, no combination that includes $\varepsilon$ 
  and any other observable will work well, 
  because the $\delta_{\alpha}(\tau,f_c)$ curves
  are invariant.
  The crucial and {\sl a priori} not so obvious conclusion is  
  that in 2$d$, in the small temperature, ideal gas regime
  the classical fields description gives at best only a qualitative description of the gas, and a description that is quantitatively correct across observables is unreachable. 
The matter of whether this is alleviated once interactions become important warrants further study. 

  \begin{figure}[t]
 \includegraphics[width=8cm,height=4.5cm]{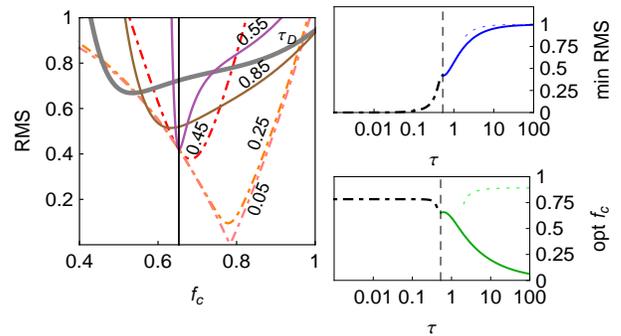}
\vspace*{-0.2cm}
  \caption{(Color online) Summary results for the 3$d$ gas.  
  The solid lines correspond to $RMS_{\varepsilon,u_{G}}$
  in the temperature region $\tau >\tau_C$
  and dot-dashed lines to $RMS_{\varepsilon,\rho_o}$
  in the region $\tau < \tau_C$.
  The left panel shows the dependence of the global error estimators 
  on cutoff $f_c$:  $\tau=\tau_D$ gray, $\tau=0.85$ brown, 
  $\tau=0.55$ purple, 
  $\tau=0.45$ red, $\tau=0.25$ orange, and $\tau=0.05$ pink.
  The black vertical line indicates the critical cutoff
  for $\tau_C$. The right panels are as in Fig.~\ref{BigRMSDim1},
  with the Bose gas critical temperature $\tau_C$ marked with
  a vertical dashed line and solutions below $\tau_C$ as dot-dashed lines.
  }
 \label{BigRMSDim3}
\end{figure}

   Fig.~\ref{BigRMSDim3} shows the results for the 3$d$ gas.
   The area above critical temperature behaves analogously 
   to low dimensions.  However, around the critical temperature,
   the $RMS$ curve narrows and the accuracy of classical fields
   becomes very sensitive to the choice of the cutoff $f_c$.
   This is related to the fluctuations
   $u_G$ growing to infinity at $\tau_C$. As such, it may be related to
   the inequivalence of the condensed ideal gas ensembles
   and may be an effect that is readily removed by finite size or interaction effects.
   In the condensed regime below $\tau_C$, the $RMS$ curve widens out again while
   classical fields rapidly become accurate 
   with $RMS<10\%$  below $\tau=0.486\, \tau_C$.

\section{Discussion}
  \label{DISC}

  Several points can be addressed on the basis of the ideal gas results,
  before considering an interacting gas.

\subsection{Nonuniform gases}

  So far, we have been fully focused on the local density approach here
  in order to obtain results that are applicable
  for general inhomogeneous cloud geometries. 

  A very convenient aspect of what we have found is that
  the best cutoff value opt$f_c$ is practically constant
  in the whole region where classical fields
  are a good description (say, $minRMS<$10\%). 
  This can be seen in Figures~\ref{BigRMSDim1} and~\ref{BigRMSDim3},
  where in this region, opt$f_c\in(0.645,0.653)$ in 1$d$
  and opt$f_c\in(0.778,0.783)$  in 3$d$.
  The best low temperature cutoff in 2$d$
  is also a constant opt$f_c=0.639$ --- see Fig.~\ref{BigRMSDim2}. 
  Even beyond this best region, the opt$f_c$ value is almost constant
  until values of $\tau\simeq1$ are reached.
  For a nonuniform gas at a temperature $T$,
  the reduced temperature scales with density as
  $\tau(\mathbf{x}) \propto 1/[n(\mathbf{x})]^{2/d}$.
  Those aspects ensure that
  the optimum cutoff for all sections of the gas
  is practically the same, 
  provided only that the bulk of the gas
  is quantum degenerate (i.e. $\tau(\mathbf{x})\lesssim1$).
  If it isn't, then the description is not accurate anyway.
  Operationally, this all means that
  the best cutoff to choose regardless of the density profile of the gas is:
  \begin{equation}
    k_c = \frac{{\rm opt}f_c}{\hbar}\,\sqrt{2\pi mk_BT}\quad;\quad \varepsilon_c = \pi ({\rm opt}f_c)^2 k_BT.
  \end{equation}
   So either we take the low temperature cutoff or it doesn't matter anyway. 
   
 The case of the uniform GCE is in fact quite well matched
 to the trapped canonical ensemble (CE) gas 
 that was mentioned in \ref{ENS},
 despite the apparently different framework of the problem. 
 This is because  the dominant central bulk of such a trapped gas
 is effectively a uniform open system in the LDA.
 The cutoffs found for the harmonically trapped canonical ensemble of the ideal gas
 based on condensate fraction distribution
 at low $\tau$ \cite{Witkowska09} correspond in our notation to values of
 \mbox{$f_c^{\rm(CE-trap)}= \{0.56, 0.72, 0.84\}$} in 1, 2, and 3$d$, respectively. 
 These are quite close to the GCE values of opt$f_c^{\rm(LDA)} = \{0.65, 0.64, 0.78\}$
 found here (Figs.~\ref{BigRMSDim1}-\ref{BigRMSDim3}). 
 This further reinforces the view that results obtained with the LDA
 are also relevant for nonuniform gases,
 even when the entire cloud
 does not have particle exchange with an environment. 
 
 A certain exception is the canonical ensemble
 in a box whose cutoffs were also studied in \cite{Witkowska09}
 and found to be much lower $f_c^{\rm(CE-box)}= \{0.30, 0.47, 0.65\}$. 
 This indicates that this is a special case
 which describes very different physics.
 The matter of which ensemble should be used to describe
 the recently achieved box potentials
 \cite{Gotlibovich14, Chomaz15, Navon15} is still open.
 If the interaction is not too strong, then
  shot-to-shot fluctuations in energy and particle number
 can be appreciable and so
 a grand canonical approach may be warranted for the whole gas
 (if one is concerned with ensemble rather than time-averaged properties).  


\subsection{Breakdown mechanism at high temperatures}

  The reason for the drop of $f_c^{u_G}$ to zero at high $\tau$ 
  provides an instructive example of how 
  the classical field description breaks down. 
  Generally the explanation comes down 
  to different statistics of 
  particle numbers $N_{\mathbf{k}}$ in the modes.
  Both the fully quantum Bose gas and the classical field
  have an exponentially decaying particle number distribution:
  \begin{equation}
    P(N_{\mathbf{k}}) \propto e^{(\mu-\hbar^2|\mathbf{k}|^2/2m)N_{\mathbf{k}}}
  \end{equation} 
  in each mode. 
  However, in the exact treatment, $N_{\mathbf{k}}$ can only take on discrete values
  $\mbox{0, 1, \dots}$, 
  while in the classical fields
  the non-integer part of the distribution is also needed.
  This peculiarity very strongly increases fluctuations especially when
  the bulk of the distribution is in this region, i.e. the mean
  number of particles is $N_{\mathbf{k}}\lesssim\mathcal{O}(1)$. 

  The above observations are transfered to $u_G$ in the following way:  
  For the exact Bose gas, the distribution is Poissonian
  when $N_{\mathbf{k}}\ll1$, giving 
  ${\rm var}[N^{\rm(id)}_{\mathbf{k}}]~=~\langle N^{\rm(id)}_{\mathbf{k}}\rangle$ for each mode, while 
  the exponential distribution in classical fields gives
  ${\rm var}[N^{(\rm(cf)}_{\mathbf{k}}]~=~\langle N^{\rm(cf)}_{\mathbf{k}}\rangle^2$. 
  Due to having independent modes, 
  $u_G=\sum_{\mathbf{k}} {\rm var}[N_{\mathbf{k}}]\ / \sum_{\mathbf{k}} \langle N_{\mathbf{k}}\rangle$,
  and in the exact treatment $u_G\to1$ directly.
  To obtain the same with classical fields,
  occupations $N_{\mathbf{k}}^{\rm(cf)}\sim1$ are necessary to make
  ${\rm var}[N_{\mathbf{k}}]\approx \langle N_{\mathbf{k}}\rangle$. 
  These are much greater than in the Bose gas. So to also simultaneously match
  overall density of the many-mode gas, the cutoff must be made much lower
  than the Bose gas momentum width $2\pi/\Lambda_T$
  so as to get the same area under the distribution of density in k-space.
  From (\ref{eq_fc}), this immediately implies $f_c^{u_G}\ll1$.
  With such a great modification of $N(\mathbf{k})$,
  correctly matching additional observables like $\varepsilon$ with classical fields
  becomes out of the question. 

 A similar breakdown can be expected whenever
 the physics is captured by low-occupied independent modes.
 For example, such discrepancies were seen between experiment
 and classical fields in the quantum Bogoliubov regime
 of the \emph{interacting} gas at very low temperatures \cite{Armijo}.

\section{Crossover to the interacting gas}
 \label{INTER}

 An obvious question is whether the ideal gas results carry over into the interacting gas.
 To address this, we have benchmarked
 the classical field description in 1$d$ system 
 with the Yang-Yang exact solution
 for the uniform interacting Bose gas \cite{Yang69} for a sequence
 of increasing interaction strengths 
 that cover the crossover
 from the ideal gas to an interaction-induced quasicondensate.  

\subsection{Procedure}
 \label{INTER_a}
 The exact values for $n=N/L$, as well as system energy $E$ in a segment of length $L$
 can be obtained via the self-consistent numerical solution of the integral equations
 given in the original Yang-Yang paper \cite{Yang69}.
 The Hellmann-Feynman theorem was used by Kheruntsyan {\it et al.}
 to obtain $g^{(2)}(0) = -\frac{1}{n^2}\left(\partial P/\partial g\right)_{\mu,T}$
 from the Yang-Yang solution for pressure $P$ \cite{Kheruntsyan03,Kheruntsyan05},
 which can be readily evaluated numerically. For the contact-interacting gas the expression for the interaction energy in the system is $E_{\rm int} = \frac{1}{2}g n^2 L g^{(2)}(0)$. From this, one obtains the kinetic energy per particle: $\varepsilon_{\rm kin}  = (E-E_{\rm int})/N$. The coarse-grained density fluctuations can also be found via $u_G = \frac{k_BT}{n}\left(\partial n/\partial \mu\right)_T$, based on the expression for ${\rm var}[N]$ in \cite{Armijo11}.

 To obtain classical field results, we generate
 ensembles of classical field realizations $\Psi(x)$
 using a Metropolis algorithm,
 in a way conceptually similar to the work of Witkowska {\it et al.} \cite{Witkowska10}
 but using grand canonical ensemble weights $e^{[\mu N(\Psi) - E(\Psi)]/k_BT}$.
 The numerical lattice is chosen to have a box of length $L$
 with periodic boundary conditions that is wide enough
 for the density and phase correlations to decay to zero before wrapping around.
 The number of points was $2^{10}$, which is easily sufficient
 for the maximum numerical lattice wavevector
 to be many times larger than the cutoffs $k_c$ imposed on the field in k-space.
 This ensures that no aliasing problems appear for the evaluation of the interaction energy term,
 as has been discussed in the context of the padded lattice in the PGPE
 and truncated Wigner methods \cite{Norrie06}.
 Classical field values for observables at a given cutoff
 are calculated using $10^4$ ensemble members. 

 For each cutoff, the observables
 are benchmarked against exact Yang-Yang values
 for systems having the same values of $T$, $g$, and density $n$ as the classical field ensemble. 
 It remains true for the interacting gas that $u_G$ and $\varepsilon_{\rm kin}$
 have the most extreme behavior among the set of observables
 that now also include the interaction energy per particle and $g^{(2)}(0)$.
 The latter two have a cutoff-dependent behavior that is somewhat similar to $u_G$.
 Hence, we continue to use the same global accuracy indicator (\ref{eqBigrms})
 as for the ideal gas, using the kinetic energy per particle $\varepsilon_{\rm kin}$
 and coarse-grained fluctuations $u_G$. 

 $minRMS$ and opt$f_c$ are obtained by fitting a function 
 to the cutoff-dependent values of $RMS(\tau,f_c)$ 
 at a given $\tau$ and $\gamma$.  
 We use the square root of a parabola 
 because it is a good candidate for describing the $f_c$-dependent
  behavior of $RMS(\tau,f_c)$ near the minimum.
 It marries the approximately linear behavior
 of $\delta_{\varepsilon}$ and $\delta_{u_G}$ in this region that
 is seen in Fig.\ref{Srms}, with the expression (\ref{eqBigrms})
 for $RMS$.  
 We use data from an $f_c$ range of about $\pm 0.05$ around the minimum.
 Error bars are obtained by splitting
 the field samples into $N_S$ smaller subensembles,
 calculating subensemble values of  $minRMS^{(i)}$ and opt$f_c^{(i)}$
 in the same way for each, and invoking the central limit theorem
 to estimate the uncertainty in the full-ensemble values
 to be $\Delta$opt$f_c=\sqrt{{\rm var}[{\rm opt}f_c^{(i)}]/N_S}$
 and  $\Delta minRMS=\sqrt{{\rm var}[minRMS^{(i)}]/N_S}$. 


\subsection{Results}

\begin{figure}[ht]
{\center
 \includegraphics[width=8cm]{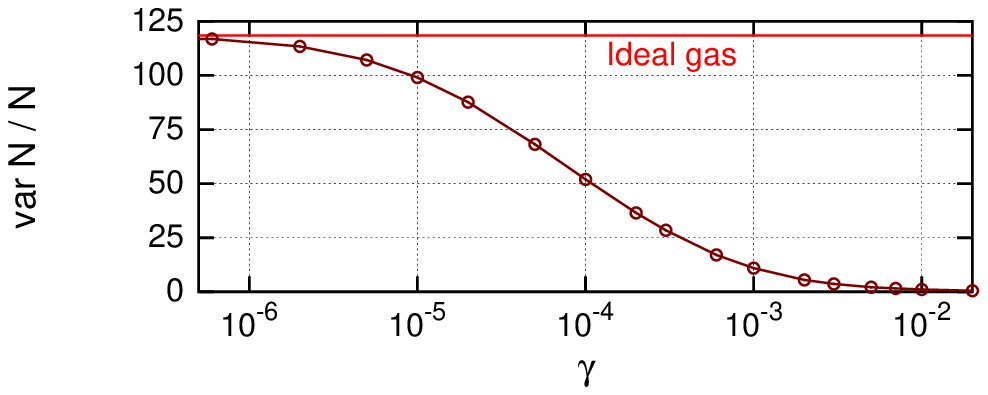}\\
 \includegraphics[width=8cm]{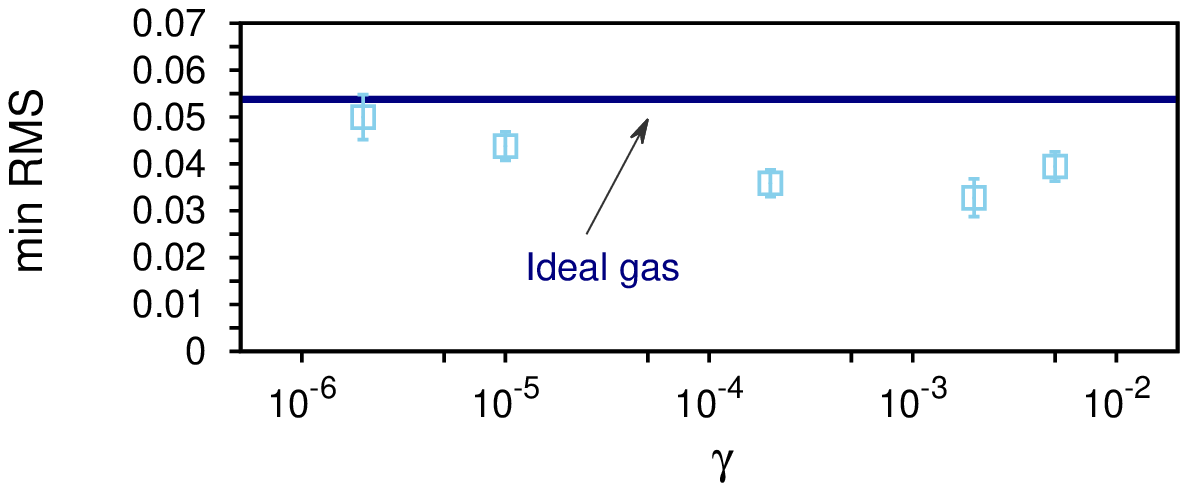}\\
 \includegraphics[width=8cm]{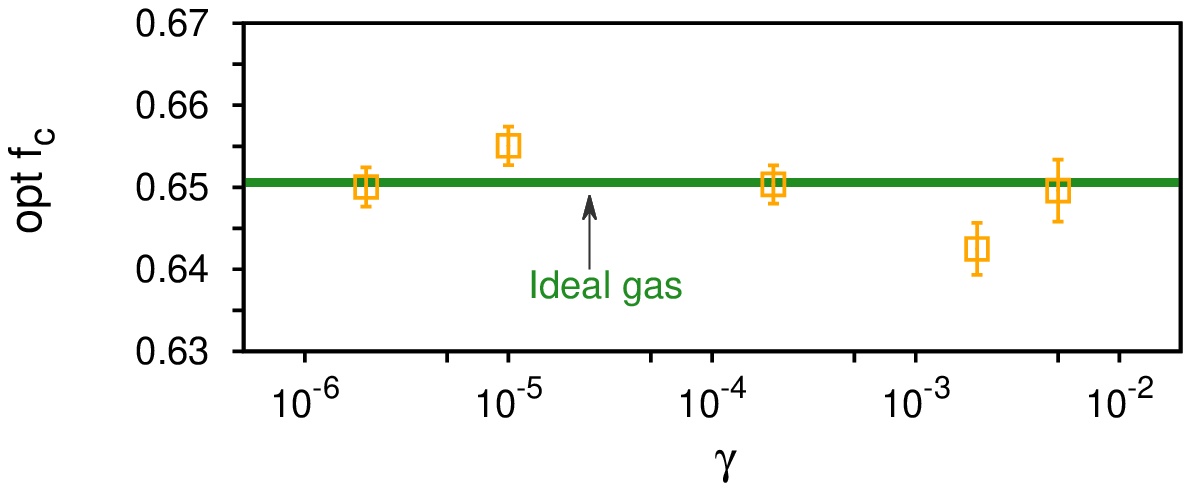}}
\vspace*{-0.2cm}
  \caption{(Color online) A preview of the situation in the $1d$ interacting gas.
   Here, $\tau=0.00159$, and $\gamma=g/n$ increases to the right.
   Top panel: the change in $u_G$, obtained from the Yang-Yang\cite{Yang69} exact solution,
   compared to the ideal gas $\gamma\to0$ value (red).
   Middle panel: minimal value of $RMS$ as in Figs.~\ref{BigRMSDim1}-\ref{BigRMSDim3}
   with 1$\sigma$ statistical error bars from an ensemble of $10^4$ samples,
   and the ideal gas value shown as the horizontal line.
   Bottom panel: corresponding optimum cutoff $f_c$ and its ideal gas value.
   One sees that while the observable $u_G$ changes by two orders of magnitude,
   the ideal gas values for cutoff and accuracy carry over onto the interacting gas.
 }\label{intFIG}
\end{figure}

 We have carried out the above benchmarking for the reduced temperature $\tau=0.00159$
 and a range of interaction strengths $\gamma=g/n$  from $2\times10^{-6}$ to $0.005$
 in the dilute interacting gas.
 These are experimentally realistic parameters.
 The local bunching $g^{(2)}(0)$ changes over this range
 from $1.976$ in the very weakly interacting limit
 to $1.02$ at $\gamma=0.005$. This indicates that we move
 from an almost perfect ideal gas on the left deep into
 the strong quasicondensate regime on the right,
 where almost all effects are dominated  by the interaction mean field.
 The coarse-grained density fluctuations change by 
 two orders of magnitude over this range,
  as plotted in the top panel of Fig.~\ref{intFIG}.
 
 The results of this foray into the interacting gas are shown in
 the other panels of Fig.~\ref{intFIG}. 
 The ideal gas values for cutoff carry over
 onto the interacting gas unchanged, to within available statistical precision. 
 The global accuracy $minRMS$ actually improves.
 One concludes then that in this regime at least
 the optimum cutoff and degree of accuracy found
 in the ideal gas applies very well to a wide swath of the interacting gas as well. 
 This is not an \emph{a priori} obvious result,
 but certainly a convenient and encouraging one
 for those who want to make calculations using classical fields.


\section{Conclusions}
\label{CONC} 
  To conclude, we have judged the goodness of
  classical fields for describing
  the ideal Bose gas in 1$d$, 2$d$, and 3$d$  
  using all the usually measured observables. 
  We have shown that $10\%$ or better accuracy for the whole set of 
  observables simultaneously is possible in 1$d$ up to temperatures of 
  $T=0.0064\, T_d$ 
  with the cutoff prescription $k_c \approx 0.65 \big(\frac{2 \pi}{\Lambda_T}\big)$ and
  in 3$d$ up to $T=0.49\, T_c$ 
  with  $k_c \approx  0.78 \, \big(\frac{2 \pi}{\Lambda_T}\big)$. 
  The essence of the matter can be captured by the indicator
  $RMS$ based on kinetic energy per particle and
  coarse grained density fluctuations,
   which are the observables that are the hardest to mutually satisfy.

  In 2$d$, we have found a surprising feature that classical fields
  remain incapable of properly describing all the observables together
  in the ideal gas even as $T \to 0$. One suspects that finite size effects
  and/or weak interactions may improve agreement here.
  The indication is that something is going on  in 2$d$ that warrants further study.
  
  When a system is correctly described with a classical ensemble of complex fields as here,
  the observation of many ``intrinsically quantum'' effects that rely on
  wave-particle duality or a discretization of the basis is ruled out.
  This includes things such as stronger-than-classical correlations,
  Heisenberg uncertainty relations, mode entanglement, EPR and 
  Bell inequality violation, antibunching, and noncommuting observables.
  All in line with the difference between classical optics on the one hand
  and quantum optics and quantum information theory on the other.
  Thus, for parameters in which the weakly interacting Bose gas
  is described by the classical field to some level of $RMS$,
  observation of the above intrinsically quantum effects
  with typical observables will also be suppressed
  to a level of the same order as $RMS$.
  Of course, large $RMS$ is not sufficient to imply quantum effects.
 
  Two results lead to optimistic conclusions for the practical application
  of classical fields to ultra cold gases.
  Firstly, the optimum cutoffs in the ideal gas are almost
  unchanged with $\tau$ in the whole region where accuracy is good.
  This means that even for a nonuniform cloud with a common global temperature,
  a single cutoff value is close to optimal in the entire degenerate region.
  This goes a long way towards pacifying
  one of the leading practical worries. 
  Secondly, our study of the crossover into the interacting gas
  in Sec.~\ref{INTER} shows that the cutoff that optimizes the ideal gas
  is also valid for a part of the interacting gas,
  including a region where the quasicondensate
  is dominated by interactions.
  The degree of accuracy seen in the ideal gas is also preserved.
  This  is a nontrivial but very encouraging result. 
  A more detailed study of the situation for the whole range
  of interaction strengths in 1$d$ is in progress and will be reported on in future.

\begin{acknowledgments}
 We would like to adress our thanks to
 Vyacheslav I. Yukalov and Matthew Davis
 for their attention and comments to us on this matter.
 This~ work was supported by the National Science Centre grant~ No.~ 2012/07/E/ST2/01389.  
\end{acknowledgments}

\end{document}